\documentclass[aps,pre,reprint,twocolumn,showpacs,floatfix,longbibliography]{revtex4-1}

\usepackage{graphicx}
\usepackage{amsmath}
\usepackage{amssymb}
\usepackage{epsf}

\begin{document}

\title{Redundancy in the information transmission in a two-step cascade}

\author{Ayan Biswas}
\email{ayanbiswas@jcbose.ac.in}
\affiliation{Department of Chemistry, Bose Institute, 93/1 A P C Road, Kolkata 700009, India}

\author{Suman K Banik}
\email[Corresponding author: ]{skbanik@jcbose.ac.in}
\affiliation{Department of Chemistry, Bose Institute, 93/1 A P C Road, Kolkata 700009, India}

\begin{abstract}
We present a stochastic framework to study signal transmission in a 
generic two-step cascade $\text{S} \rightarrow \text{X} \rightarrow \text{Y}$. 
Starting from a set of Langevin equations obeying Gaussian noise 
processes we calculate the variance and covariance while considering 
both linear and nonlinear production terms for different biochemical 
species of the cascade. These quantities are then used to calculate 
the net synergy within the purview of partial information decomposition. 
We show that redundancy in information transmission is essentially an
important consequence of Markovian property of the two-step cascade
motif. We also show that redundancy increases fidelity of the signalling 
pathway.
\end{abstract}


\date{\today}

\maketitle


\section{Introduction}

A living system sustains in the diverse and continuously changing environment. 
In order to respond to the changes made in the surroundings, every living 
species developed complex signal transmission networks over the
evolutionary time scale \cite{Alon2006,Alon2007}. The main purpose of these 
networks is
to transmit the extracellular changes reliably and efficiently to the cell. In
addition, these networks take care of different biochemical changes that are taking
place within the cell. A typical signalling cascade comprises of one or 
several components of biochemical origin. The interactions between these 
components are probabilistic in nature, thus giving rise to stochastic kinetics. 
One of the tools to figure out the signal transmission mechanism in a fluctuating 
environment is information theory \cite{Shannon1948,Cover1991}. The 
formalism of information theory provides a quantification of information 
transfer between the source (signal) and the target (response). The 
measure of information transmission is characterized by mutual information 
(MI), that quantifies the common information content of the source and the 
target. Moreover, information theory provides a measure of fidelity of the 
signalling pathway \cite{Borst1999,Mitra2001,Ziv2007,Tostevin2010,Lestas2010,Cheong2011,
Tkacik2011,Bowsher2013,Maity2015,Mahon2014,Tsimring2014}.

The notion of MI is conceptualized as the intersection of the entropy spaces 
of two stochastic variables \cite{Shannon1948,Cover1991}. Hence, MI signifies 
the average reduction in the uncertainty of prediction of one random variable 
when knowledge of another random variable is available. However, this 
information theoretic measure is symmetric in its argument random variables 
signifying the `mutual' attribution linked to it. For a generic two-step cascade
(TSC) 
$\text{S} \rightarrow \text{X} \rightarrow \text{Y}$, although, MI among three 
variables is an ill defined concept, its usage can be validated if one considers 
$I(s; x,y)$ to be the MI that the source species S  shares with the pair of target 
species X and Y. It is thus interesting to investigate whether the three variable 
MI can be utilised in such a way to shed light on various types of informational
relationship among the three stochastic variables ($s$, $x$, and $y$). 
One prevalent approach in this respect is known as partial information 
decomposition (PID) \cite{Williams2010,Barrett2015}. 
Following the formalism of PID one can define an information theoretic measure, 
the net synergy $\Delta I(s; x,y)$, in terms of two and three variable MI-s
\cite{Schneidman2003,Williams2010,Barrett2015,Hansen2015,Maity2016}
\begin{equation}
\label{eq1}
\Delta I(s; x,y) = I(s; x,y) - I(s; x) - I(s; y).
\end{equation}

\noindent
At this point, one should take note of the fact that the net synergy can be both 
positive ($\Delta I > 0$) and negative ($\Delta I < 0$) valued. Information 
theoretically, 
positive net synergy implies synergistic aspect of the target variables is prevalent 
over the extent of redundant character. Negative net synergy conveys precisely the 
opposite implication. Positive net synergy indicates the information shared 
between the source S and the targets (X, Y) taken together as a single 
target is more than the sum of the information shared between S and the targets 
X, Y considered individually in turn. Negative net synergy indicates separately the 
targets share more information with the source than while considered together as a 
single target. Zero net synergy ($\Delta I = 0$) is an interesting case to mention 
as it means that 
the targets X and Y are informationally independent of each other since the 
information shared between the source and the targets does not depend on 
whether one takes the targets individually or together as a single target to compute 
different MI terms. As per the definition of the net synergy stated in 
Eq.~(\ref{eq1}), it is maximum while the total information is synergistic and
assumes a minimum value when the total information is purely redundant in 
nature. In this study, we do not quantify the separate identification of synergy 
and redundancy. Rather, we conceive the quantity of interest here as redundant 
synergy. Information theoretically, the net synergy serves as a quantification marker 
of information independence between the random variables often characterized 
as source and target variables \cite{Schneidman2003}. Although, the tags of 
source(s) and target(s) are not strict enough in information theoretic sense but 
while dealing with cascades or directed networks, in general, such classifications 
can be done suitably as the phenomenology or experimental realities demand.

Information theoretic ideas originated in Shannon's classic work 
\cite{Shannon1948} 
had formerly made an impressive impact in the field of biology through 
experimental neuroscience. In that domain, researchers had successfully 
used the technique to quantify the extent of information about the stimulus 
encoded in a neural response \cite{Borst1999}. Earlier work regarding 
efficient communication systems has dealt with the capacity of nonlinear 
information channel considering the physical limits \cite{Mitra2001}. 
Investigations on maximization of MI in biochemical 
networks of different topologies have also shown the robustness of
information transduction even with considerably low molecule numbers
\cite{Ziv2007}. Recent theoretical papers dealing with MI has done so in 
the Gaussian framework given the benefit of exactness of calculated MI
\cite{Tostevin2010}. It should be also noted for a Gaussian system, MI 
becomes equal to the channel capacity \cite{Tostevin2010}.
The same work also draws our attention to the fact that application of linear
noise approximation applied in a network of interest is compatible with 
the Gaussian model. This line of approach has been taken care of in 
the present work. Earlier works suggest quantitative agreement between 
steady state solutions obtained through linear noise approximation and 
simulations carried with Gillespie algorithm even when biochemical molecules 
have low copy numbers ($\sim 10$) \cite{Ziv2007,Bruggeman2009}. 
Consideration of low copy numbers in the dynamics shows transmission 
of more than 1 bit of information.
Our theoretical calculation reveals that not only the individual two and 
three variable MI-s but also the net synergy which acts as the predictor of 
informational independence can have magnitudes of greater than 1 bit 
even when signalling molecule has low copy number (see Sec.~III).

The idea of synergy has been explored in the previous work of Schneidman 
et al. \cite{Schneidman2003} in the context of independence in neural coding. 
Though the same concept is reiterated in Ref.~\cite{Barrett2015}, there it 
comes with a different name, the net synergy. The idea of net synergy arises 
as a product of PID for Gaussian variables. Beyond the limits of neurophysiological 
domain, information theory has also got prominence in analysis of gene 
regulatory networks. In this connection it is important to mention earlier 
set of works where optimization of information flow in genetic networks 
have been addressed \cite{Tkacik2009,Walczak2010}. 
Other insightful information theoretic treatments of genetic regulation can be 
found in Refs.~\cite{Tkacik2008,Tkacik2008a} and in a review article 
by Tka{\v c}ik and Walczak \cite{Tkacik2011}. Bruggeman et al. have
successfully applied metabolic control analysis along with linear noise 
approximation to analyze how feedback mechanism and separation of
time scale can affect noise flow in molecular networks and the influence
of network architecture on noise \cite{Bruggeman2009}.
Usage of nonlinear regulatory functions in 
Refs.~\cite{Tkacik2009,Walczak2010,Ronde2012} made the analysis close 
to the biological realities prompting us to study simplified linear model along 
with the nonlinear one to compare the fluctuations level of the system under 
both scenarios.
In Ref~\cite{Tkacik2009}, the authors have studied a network where multiple 
target genes receive signalling from a single input. It is noted that those target 
genes turn on at successively higher levels of concentration of the input signal. 
Consequently, it is pointed out that there happens to be redundancy among the 
signals conveyed. Unlike the treatment rendered in 
Refs.~\cite{Tkacik2009,Walczak2010}, we kept the input concentration, along
with other species concentrations, fixed at steady state to extract optimal 
functionality out of the constituents of the TSC motif and tuned the degradation 
rate of the input signal \cite{Alon2006,Maity2015}.

The first focal theme of our work is to investigate redundancy in the information 
transduction process using the generalized mathematical measure of the net 
synergy (see Eq.~(\ref{eq1})) along with the adoption of a motif based approach.
In this connection, it is to be noted that noise has an important role to play in the
information transmission machinery. An excellent review \cite{Tsimring2014} has 
covered the ubiquitous functionalities of noise in several biological phenomena.
Ref.~\cite{Lestas2010} deals with suppression of the noise by feedback mechanisms 
within the physical limits. Earlier work \cite{Walczak2010} also hinted at the tradeoff 
between redundancy reduction and noise reduction.
The experimental work \cite{Cheong2011} has also strengthened this 
viewpoint by pointing out that TNF signalling pathway can overcome noise 
induced limitations on transmitting information through adoption of complex 
signalling networks which by construction incorporate redundancies. Our 
second focal theme in the present work has been analyzing this nontrivial 
connection between redundancy and noise in a quantitative manner to shed 
light on the fidelity of the motif. In this connection, Bowsher et al. suggests 
that the signal-to-noise ratio (SNR) to be an able quantifier of fidelity 
\cite{Bowsher2013} which we have adopted. 
Relevant work that connects feedback and the fidelity with which information
is transmitted has been also reported in the existing literature \cite{Ronde2010}.
The PID route to study information processing also reveals the Markovian 
structure of the TSC motif and shows the motif's adherence to data processing 
inequality (DPI) \cite{Cover1991}.

To understand the redundancy in the information transmission in biochemical
processes we undertake the kinetics associated with a generic TSC Motif
$\text{S} \rightarrow \text{X} \rightarrow \text{Y}$.
The dynamics related to the TSC motif could be linear or nonlinear in nature and
provides analytical solution at steady state within the purview of Gaussian
noise processes. In this connection, it is important to mention theoretical analysis
performed on different biochemical motifs obeying Gaussian noise processes 
\cite{Bialek2005,Ziv2007,Ronde2010,Tanase2006,Warren2006,Bruggeman2009,
Tostevin2010,Maity2014,Maity2015,Grima2015,Maity2016}.
In this set of works, the theoretical analysis was performed using linear noise 
approximation \cite{Kampen2007,Elf2003,Gillespie2000} that provides exact 
expressions upto second moments.
Recent theoretical development \cite{Grima2015} shows that linear noise 
approximation is not only limited for high copy number conditions but can also 
be exact up to second moments for some systems with second-order reactions.
In this case, it is found that the fluctuations associated with at least one
of the species participating in each of the second-order reactions are 
Poissonian and uncorrelated with the fluctuations of the other species.

%
\section{The Model}

The set of Langevin equations governing the dynamics of a TSC motif can 
be written as,
\begin{eqnarray}
\label{eq2}
\frac{ds}{dt} & = & f_s (s) - \mu_s s + \xi_s(t), \\
\label{eq3}
\frac{dx}{dt} & = & f_x (s,x) - \mu_x x + \xi_x(t), \\
\label{eq4}
\frac{dy}{dt} & = & f_y (s,x,y) - \mu_y y + \xi_y(t),
\end{eqnarray}

\noindent
where  $s$, $x$ and $y$ are copy numbers of the species S, X and Y,
respectively, expressed in molecules/$V$ with $V$ being the unit 
cellular volume.
Here $f_i$-s and $\mu_i$-s ($i=s, x, y$) are the 
synthesis and degradation rates of the components S, X and Y, 
respectively. Depending on the nature of interaction, the synthesis terms could 
be linear or nonlinear in nature. The noise terms $\xi_i (t)$ are independent and
Gaussian-distributed with properties $\langle \xi_i (t) \rangle = 0$ and 
$\langle \xi_i (t) \xi_j (t') \rangle =  \langle | \xi_i |^2 \rangle \delta_{ij} \delta (t-t')$,
where 
$ \langle | \xi_i |^2 \rangle 
= \langle f_i \rangle + \mu_i \langle i \rangle 
= 2 \mu_i \langle i \rangle$ for $i=s, x, y$
\cite{Elf2003,Swain2004,Paulsson2004,Tanase2006,Warren2006,Kampen2007,Mehta2008,Ronde2010}.
The quantities $\langle i \rangle = \langle s \rangle, \langle x \rangle, 
\langle y \rangle$ are the ensemble average of the respective species 
determined at steady state \cite{Ronde2012}.
We note that the usage of constant noise intensity evaluated at steady state,
in the present analysis, is an approximation that makes the following analytical 
calculation tractable.
We now expand Eqs.~(\ref{eq2}-\ref{eq4}) around steady state
$\delta z(t) = z(t) - \langle z \rangle$, where $\langle z \rangle$ is the
average population of $z$ at steady state and obtain,
\begin{equation}
\label{eq5}
\frac{d \delta \mathbf{A}}{dt} =
\mathbf{J}_{A = \langle A \rangle} \delta \mathbf{A} (t) + \mathbf{\Theta} (t).
\end{equation}

\noindent
Here $\delta \mathbf{A}$ and $\mathbf{\Theta}$ are the fluctuations matrix 
and the noise matrix, respectively, with
\begin{eqnarray*}
\delta \mathbf{A} = \left (
\begin{array}{c}
\delta s \\
\delta x \\
\delta y
\end{array} 
\right ) 
\; {\rm and} \;
\mathbf{\Theta} = \left (
\begin{array}{c}
\xi_s \\
\xi_x \\
\xi_y
\end{array} 
\right ).
\end{eqnarray*}

\noindent $\mathbf{J}$ is the Jacobian matrix evaluated at steady state
\begin{widetext}
\begin{eqnarray*}
\mathbf{J} = \left (
\begin{array}{ccc}
f^{\prime}_{s,s} (\langle s \rangle) -\mu_s & 0 & 0 \\
f^{\prime}_{x,s} (\langle s \rangle, \langle x \rangle) & 
f^{\prime}_{x,x} (\langle s \rangle, \langle x \rangle) - \mu_x & 0 \\
f^{\prime}_{y,s} (\langle s \rangle, \langle x \rangle, \langle y \rangle) & 
f^{\prime}_{y,x} (\langle s \rangle, \langle x \rangle, \langle y \rangle) & 
f^{\prime}_{y,y} (\langle s \rangle, \langle x \rangle, \langle y \rangle) - \mu_y
\end{array} 
\right ),
\end{eqnarray*}
\end{widetext}

\noindent
where $f^{\prime}_{s,s} (\langle s \rangle)$ implies $f_s$ has been differentiated
with respect to $s$ and is evaluated at $s = \langle s \rangle$, and so on.
To calculate the variance and covariance of different species of the TSC motif, 
we use the Lyapunov equation at steady state
\cite{Keizer1987,Kampen2007,Elf2003,Paulsson2004,Paulsson2005}
\begin{equation}
\label{eq6}
\mathbf{J} \mathbf{\Sigma} + \mathbf{\Sigma} \mathbf{J}^T 
+ \mathbf{D} = 0.
\end{equation}

\noindent
Here, $\mathbf{\Sigma}$ is the covariance matrix and 
$\mathbf{D} = \langle \mathbf{\Theta} \mathbf{\Theta}^T \rangle$ is the
diffusion matrix due to different noise strengths.
The notation $\langle \cdots \rangle$ stands for ensemble average
at steady state and $T$ is the transpose of a matrix. For linear interaction
\cite{Ronde2010,Ronde2012,Maity2015}
\begin{eqnarray*}
f_s = k_s, f_x = k_x s \; {\rm and} \; f_y = k_y x, 
\end{eqnarray*}

\noindent 
solution of Eq.~(\ref{eq6}) provides analytical expressions of variance 
and covariance associated with $s$, $x$ and $y$
\begin{eqnarray*}
\Sigma (s) & =& \langle s \rangle,
\Sigma (s,x) = \frac{k_x \langle s \rangle}{\mu_s + \mu_x}, \\
\Sigma (s,y) &=& \frac{k_y k_x \langle s \rangle}{(\mu_s + \mu_x) (\mu_s + \mu_y)}, \\
\Sigma (x) & =& \langle x \rangle + \frac{k_x^2 \langle s \rangle}{\mu_x (\mu_s + \mu_x)}, \\
\Sigma (x,y) &=& \frac{k_y}{\mu_x + \mu_y} \Sigma (x) 
+ \frac{k_x}{\mu_x + \mu_y} \Sigma (s,y), \\
\Sigma (y) &=& \langle y \rangle + \frac{k_y}{\mu_y} \Sigma (x,y).
\end{eqnarray*}

\noindent
Similarly, for nonlinear interaction 
\cite{Ziv2007,Bintu2005b,Bialek2008,Tkacik2008a,Tkacik2008b}
\begin{eqnarray*}
f_s = k_s, f_x = k_x \frac{s^n}{K_1^n + s^n} \; {\rm and} \; 
f_y = k_y \frac{x^n}{K_2^n + x^n},
\end{eqnarray*}

\noindent we have from Eq.~(\ref{eq6})
\begin{eqnarray*}
\Sigma (s) & =& \langle s \rangle,
\Sigma (s,x) = \frac{nk_x K_1^n \langle s \rangle^n}{(K_1^n + \langle s \rangle^n)^2 (\mu_s + \mu_x)}, \\
\Sigma (s,y) & = & \frac{n k_y K_2^n \langle x \rangle^{n-1}\Sigma (s,x)}{(K_2^n + \langle x \rangle^n)^2 (\mu_s + \mu_y)}, \\
\Sigma (x) & =& \langle x \rangle + \frac{nk_x K_1^n \langle s \rangle^{n-1}\Sigma (s,x)}{(K_1^n + \langle s \rangle^n)^2 \mu_x}, \\
\Sigma (x,y) &=& \frac{n k_y K_2^n \langle x \rangle^{n-1}\Sigma (x)}{(K_2^n + \langle x \rangle^n)^2 (\mu_x + \mu_y)}
+ \frac{n k_x K_1^n \langle s \rangle^{n-1}\Sigma (s,y)}{(K_1^n + \langle s \rangle^n)^2 (\mu_x + \mu_y)}, \\
\Sigma (y) &=& \langle y \rangle + \frac{n k_y K_2^n \langle x \rangle^{n-1}\Sigma(x,y)}{(K_2^n + \langle x \rangle^n)^2 \mu_y}.
\end{eqnarray*}

We note that, in the above set of expressions of different variance and 
covariance evaluated at steady state (for both linear and nonlinear 
interactions), we have approximated $s$, $x$ and $y$ by $\langle s \rangle$, 
$\langle x \rangle$ and $\langle y \rangle$, respectively 
\cite{Maity2015,Ronde2010}. We now quantify the MI-s associated with 
the signalling cascade $\text{S} \rightarrow \text{X} \rightarrow \text{Y}$. 
For $s$, $x$ and $y$ assumed to be Gaussian random variables one 
can express the net synergy associated with TSC motif as follows
\cite{Barrett2015}
\begin{eqnarray}
\label{eq7}
\Delta I(s;x,y) & = & \frac{1}{2} 
\left [
\log_2 \left ( \frac{\det\Sigma(s)}{\det\Sigma (s|x,y)} \right ) \right. \nonumber \\
&& - \log_2 \left ( \frac{\det\Sigma(s)}{\det\Sigma(s|x)} \right )  \nonumber \\
&& \left. - \log_2 \left ( \frac{\det\Sigma(s)}{\det\Sigma(s|y)} \right )
\right ] .
\end{eqnarray}

\noindent
The usage of base 2 in the logarithm functions suggests that the net synergy is 
calculated in the units of `bits'. The first, second and the third term on the right 
hand side of Eq.~(\ref{eq7}) corresponds to $I(s; x,y)$, $I(s; x)$ and $I(s; y)$,
respectively. The definitions of various conditional variances used in 
Eq.~(\ref{eq7}) are \cite{Barrett2015}
\begin{eqnarray}
\label{eq8}
\Sigma(s|x) & =: & \Sigma(s)-\Sigma(s,x)(\Sigma(x))^{-1}\Sigma(x,s), \\
\label{eq9}
\Sigma(s|y) & =: & \Sigma(s)-\Sigma(s,y)(\Sigma(y))^{-1}\Sigma(y,s), \\
\label{eq10}
\Sigma(s|x,y) & =: & \Sigma(s) 
- \left( \begin{array}{ccc} \Sigma(s,x) & \Sigma(s,y) \end{array} \right) \nonumber \\
&& \times
\left( \begin{array}{ccc} \Sigma(x)& \Sigma(x,y) \\ 
\Sigma(y,x)&\Sigma(y) \end{array} \right)^{-1} 
\left( \begin{array}{ccc} \Sigma(s,x) \\ 
\Sigma(s,y) \end{array} \right). \nonumber \\
\end{eqnarray}

\noindent 
The expression of $\Sigma(s|x,y)$ after completing the matrix multiplication 
yields
\begin{eqnarray}
\label{eq11}
\Sigma(s|x,y) & = &\Sigma(s) - (1/{\cal D})
[ \Sigma(y) \Sigma^2(s,x) \nonumber \\
&& - 2\Sigma(s,x)\Sigma(s,y)\Sigma(x,y) \nonumber \\
&& + \Sigma(x) \Sigma^2(s,y) ],
\end{eqnarray}

\noindent
with ${\cal D}=:\Sigma(x)\Sigma(y) - \Sigma^2(x,y)$.


\section{Results and Discussion}

The analysis presented in the previous section provides a recipe
of calculation of the net synergy in a TSC motif within the purview of
linear noise approximation. The associated variance and 
covariance expressions for linear and nonlinear interactions are
general within the approximation scheme. To check the validity 
of our theoretical expressions we also carry out numerical simulation 
following stochastic simulation algorithm \cite{Gillespie1976,Gillespie1977} 
while considering the kinetics associated with the linear and 
nonlinear interactions. The chemical reactions and the propensities 
used in the simulation are given in Table~I. The numerical simulation 
also provides an understanding of the contribution of the rate parameters 
on the general expression of the net synergy.
While calculating the net synergy $\Delta I(s; x,y)$, various mutual
information ($I(s;x,y)$, $I(s;x)$ and $I(s;y)$) and SNR we use
$\langle s \rangle = 10$, $\langle x \rangle = 100$ and $\langle y \rangle = 100$
for both linear and nonlinear interactions. This helps us to compare the
behaviour of the motif at steady state for different interactions as the
level of system components at steady state performs the optimal 
function \cite{Alon2006}.
To keep the population of S constant at steady state we use the relation 
$k_s = \mu_s \langle s \rangle$ (see Eq.~(\ref{eq2})). Due to this 
relation, if one varies $\mu_s$ as an independent parameter, value of 
$k_s$ varies accordingly for fixed $\langle s \rangle$.  In case of linear 
interaction, we adopt similar strategy by using 
$k_x = \mu_x \langle x \rangle/\langle s \rangle$ and 
$k_y = \mu_y \langle y \rangle/\langle x \rangle$ to keep the copy numbers 
of X and Y fixed (see Eqs.~(\ref{eq3}-\ref{eq4})).  For nonlinear interaction 
we employ 
$k_x = \mu_x \langle x \rangle (K_1 + \langle s \rangle)/\langle s \rangle$,
$k_y = \mu_y \langle y \rangle (K_2 + \langle x \rangle)/\langle x \rangle$,
for $n = 1$.
At this point it is important to note that the above expressions of $k_s$, 
$k_x$ and $k_y$ (for linear and nonlinear interactions) are used in the 
propensity functions for simulation, as well as in the theoretical
calculation so that the fixed copy numbers of $s$, $x$ and $y$ are
maintained at steady state.


\begin{table}
\caption{
Table of chemical reactions and propensities for TSC motif.
The first six reactions are for linear interaction and the rest
are for nonlinear interaction ($n=1$). Here, S, X and Y stand 
for chemical species and $s$, $x$ and $y$ represent copy 
numbers of the respective species expressed in molecules/$V$
with $V$ being the unit cellular volume. The unit of corresponding
rate constants is min$^{-1}$.
}
\begin{ruledtabular}
\begin{tabular}{lll}
& Reaction & Propensity \\
\hline
Synthesis of S  & $\phi \rightarrow$ S & $k_s$ \\
Degradation of S  & S $\rightarrow \phi$ & $\mu_s s$ \\
S mediated synthesis of X  & S $\rightarrow$ S + X & $k_x s$ \\
Degradation of X  & X $\rightarrow \phi$ & $\mu_x x$ \\
X mediated synthesis of Y  & X $\rightarrow$ X + Y & $k_y x$ \\
Degradation of Y  & Y $\rightarrow \phi$ & $\mu_y y$ \\
\hline
Synthesis of S  & $\phi \rightarrow$ S & $k_s$ \\
Degradation of S  & S $\rightarrow \phi$ & $\mu_s s$ \\
S mediated synthesis of X  & S $\rightarrow$ S + X & $k_x \frac{s}{K_1+s}$ \\
Degradation of X  & X $\rightarrow \phi$ & $\mu_x x$ \\
X mediated synthesis of Y  & X $\rightarrow$ X + Y & $k_y \frac{x}{K_2+x}$ \\
Degradation of Y  & Y $\rightarrow \phi$ & $\mu_y y$ \\
\end{tabular}
\end{ruledtabular}
\end{table}

\begin{figure}[!t]
\begin{center}
\includegraphics[width=1.0\columnwidth,angle=0]{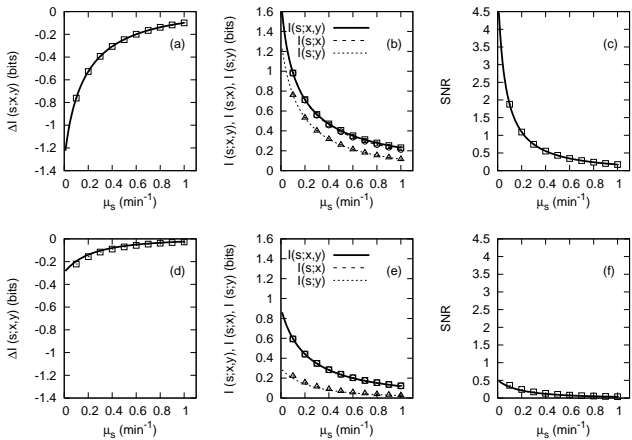} 
\end{center}
\caption{The net synergy [(a), (d)], mutual information [(b), (e)] and SNR 
[(c), (f)] profiles as functions of $\mu_s$. (a)-(c) are for linear interaction 
and (d)-(f) are for nonlinear interaction.
The lines are theoretical results and the symbols are generated 
from stochastic simulation. The simulation results are average of $10^5$ 
trajectories.
The parameters common to both linear and nonlinear interactions are
$\langle s \rangle = 10$, $\langle x \rangle = 100$, $\langle y \rangle = 100$,
$\mu_x = 0.5$ min$^{-1}$, $\mu_y = 5$ min$^{-1}$ and 
$k_s = \mu_s \langle s \rangle$.
For linear interaction 
$k_x = \mu_x \langle x \rangle/\langle s \rangle$ and 
$k_y = \mu_y \langle y \rangle/\langle x \rangle$.
For nonlinear interaction
$k_x = \mu_x \langle x \rangle (K_1 + \langle s \rangle)/\langle s \rangle$,
$k_y = \mu_y \langle y \rangle (K_2 + \langle x \rangle)/\langle x \rangle$,
$K_1 = 10$, $K_2 = 100$ and $n = 1$
(see main text).
}
\label{fig1}
\end{figure}

In Fig.~\ref{fig1}, we show the profiles of the net synergy $\Delta I(s; x,y)$, 
various mutual information ($I(s;x,y)$, $I(s;x)$ and $I(s;y)$) and SNR 
as functions of input relaxation rate constant $\mu_s$ for both linear
and nonlinear interactions ($n=1$). The lines are due to theoretical calculation 
and the symbols are generated from stochastic simulation 
\cite{Gillespie1976,Gillespie1977}. In Fig.~\ref{fig1}(a) and \ref{fig1}(d), 
both the net 
synergy profiles grow hyperbolically as $\mu_s$ is increased and 
move towards $\Delta I = 0$. For low value of $\mu_s$ the domain 
of redundancy significantly decreases as the interactions between 
the different system 
components changes from linear (see Fig.~\ref{fig1}(a)) to nonlinear 
(see Fig.~\ref{fig1}(d)). In case of nonlinear interaction with $n=1$, 
fluctuations associated with the production of X decrease as 
$f_{x,s}^{\prime} ({\rm linear}) >  f_{x,s}^{\prime} ({\rm nonlinear})$ 
for fixed copy number and parameter set. Similar relation holds good
for Y also, i.e., 
$f_{y,x}^{\prime} ({\rm linear}) >  f_{y,x}^{\prime} ({\rm nonlinear})$.
These effects together lower the magnitudes of variance and covariance 
associated with X and Y thus lowering the magnitudes of  different MI-s. 
As a result, the domain of redundancy decreases.
At this point it is important to mention that
with the adoption of $\langle s \rangle = 10$ the numerical results 
obtained from stochastic simulation algorithm agree with the analytical 
results derived using linear noise approximation, and is in agreement 
with the analysis presented earlier \cite{Ziv2007,Bruggeman2009}.



As mentioned earlier, we are only interested in the difference between 
synergy and redundancy since their individual values can not be determined 
within the purview of PID adopted in the present work.  If at all one tries 
to infer synergy from the net synergy, one may consider the net synergy 
as redundant synergy of some kind \cite{Williams2010}. 
To understand the nature of the net synergy we look at the profiles of its 
three ingredients, viz., $I(s;x,y)$, $I(s;x)$ and $I(s;y)$ as functions of 
$\mu_s$. In Fig.~\ref{fig1}(b) and \ref{fig1}(e), $I(s;x,y)$ and $I(s;x)$ are 
nearly equal while $I(s;y)$ assumes a lower value compared to the other 
two expressions of MI. This result suggests that in the TSC motif the 
relevant information is lost and is never regained while getting transduced 
from the source to the output. This loss cannot be undone or compensated 
by any kind of manipulation in the signal transduction pathway. From 
the information theoretic point of view, this is a consequence of DPI
\cite{Cover1991}. Within the framework of 
Markov chain property we have $I(s;x,y) = I(s;x)$ and $I(s;x) \geqslant I(s;y)$ 
where the inequality expression is due to DPI.  From 
Fig.~\ref{fig1}(b) and \ref{fig1}(e) it is clear that $I(s;x,y) \approx I(s;x)$ 
and $I(s;x) > I(s;y)$. Recalling the expression of the net synergy, we 
notice that the first two terms ($I(s;x,y)$ and $I(s;x)$) nearly cancel each 
other and we are left with $\Delta I (s;x,y) \approx - I(s;y)$ that generates 
the net synergy profiles shown in Fig.~\ref{fig1}. 
Introduction of nonlinearity reduces the contribution of $I(s;y)$ (the dotted 
line with triangles in Fig.~\ref{fig1}(e)) in the expression of the net synergy. 
Whereas, for linear interaction the same has a relatively high contribution 
(the dotted line with triangles in Fig.~\ref{fig1}(b)). 
Linear interaction has the ability to achieve more than 1 bit of two and 
three variable MI-s as well as the net synergy while nonlinear interaction 
compromises on these magnitudes.
This in turn supports our argument of lowering of redundancy due to lesser 
contribution of $I(s;y)$ for nonlinear interaction.

The net synergy profiles and the contributions of individual MI-s suggest that
with an increase of $\mu_s$ the common or redundant information between
X and Y decreases. A consequence of redundancy gets reflected through 
SNR defined as 
$\Sigma^2(s,y)/(\Sigma (s) \Sigma (y) - \Sigma^2 (s,y))$.
The profiles of SNR are shown in Fig.~\ref{fig1}(c) and \ref{fig1}(f). To keep 
stock of the SNR profiles, one should remember that information flow and 
noise propagation are by default antagonistic to each other. As redundant 
information increases, there are lesser chances that the system would lose 
any valuable information since even if some amount of information gets 
corrupted by noise along the signalling pathway, there are possible 
replacements of lost information due to its redundancy property. In this way, 
redundancy empowers fidelity (or SNR) of the signalling pathway. The SNR 
profiles shown in Fig.~\ref{fig1}(c) and \ref{fig1}(f) thus show opposite trend 
as compared to that of the net synergy.


\begin{figure}[!t]
\begin{center}
\includegraphics[width=1.0\columnwidth,angle=0]{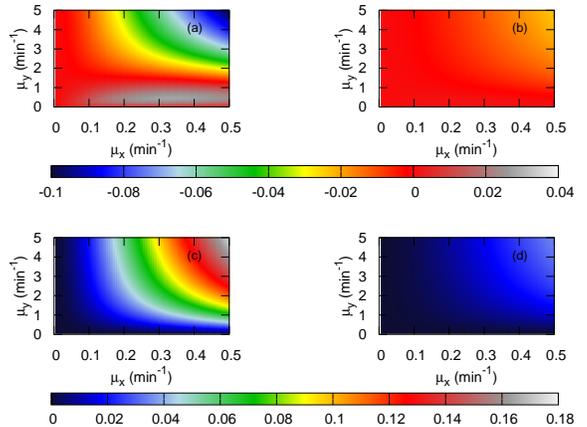} 
\end{center}
\caption{(color online) Two-dimensional maps of the net synergy 
(in bits) [(a), (b)] 
and SNR [(c), (d)] as functions of $\mu_x$ and $\mu_y$ for $\mu_s = 1$
min$^{-1}$. (a), (c) are for linear interaction and (b), (d) are for nonlinear 
interaction with $n=1$. The parameters are same as in Fig.~\ref{fig1} except
$\mu_s = 1$ min$^{-1}$. The maps are generated using theoretical 
expressions.
}
\label{fig2}
\end{figure}

From the nature of the net synergy profiles shown in Fig.~\ref{fig1}, one is 
left with the impression that the net synergy is constrained only in the
negative domain. To explore further we scan the full parameter range
of $\mu_x$ and $\mu_y$ for $\mu_s=1$ min$^{-1}$. The resultant
two-dimensional maps of the net synergy and SNR are shown in 
Fig.~\ref{fig2}. Fig.~\ref{fig2}(a) suggests that for linear interaction 
a region exists with $\Delta I > 0$. Similar trend is observed
for nonlinear interaction in Fig.~\ref{fig2}(b). Since the values of individual
MI-s are always $\geqslant 0$ \cite{Cover1991}, the positive value of the
net synergy suggests that $I(s;x,y) > I(s;x)+I(s;y)$
(see the expression of the net synergy given in Eq.~(\ref{eq1})). As 
expected, opposite trend is observed in SNR (Fig.~\ref{fig2}(c) and \ref{fig2}(d)). 
For $\Delta I > 0$ and $\Delta I < 0$ we have low and high SNR, respectively.


\section{Conclusion}

To summarize, we have investigated how different constituents of a 
generic TSC motif are related to each other in information theoretic 
sense. To investigate such mutual dependencies, we explored the 
concept of net synergy, an essential information theoretic measure 
due to the formalism of partial information decomposition. The two 
variable and three variable mutual information quantities have been 
computed in terms of variance and covariance of the Gaussian random 
variables representing the components of the TSC motif. 
Calculations presented in this work e.g., expressions of variance and 
covariance are general within the Gaussian model adopted and we 
have chosen biologically relevant parameter sets to explore interesting 
patterns of the quantities of interest.
To be specific, in this study, we have tuned the signal by changing the 
degradation rate of the source ($\mu_s$) and have quantified the three 
MI-s, the net synergy and the SNR for linear and nonlinear interactions. 
Our results show that $I(s;x,y)$ and $I(s;x)$ are nearly equal. As a 
consequence of Markov chain property, the net synergy $\Delta I(s;x,y)$ 
picks up contribution mostly from $I(s;y)$, thus showing redundancy.

We have compared simplified linear model and realistic nonlinear
model side by side and observed that introduction of nonlinearity lowers 
the magnitudes of MI-s and the net synergy which in linear case can have 
values greater than 1 bit.
Based on our findings, we argue that redundancy can increase the 
fidelity of the TSC motif as redundant information enhances SNR
in the system. 
It is to be noted that compared to the idealistic linear regulation case, 
nonlinear regulation enters the system with a point of disadvantage i.e.,
the reduced fidelity with which input signal can be relayed to 
the output response.
We further make a thorough scan of the parameter space and notice
a region of synergy where $I(s;x,y) > I(s;x)+I(s;y)$.
The quantitative analysis of redundancy in a biological motif
through generalized measure of the net synergy using realistic
regulatory model and parameters and establishing quantitative
relationship between redundancy and fidelity are the key points
of our study which to the best of our knowledge, are new additions
to the existing literature. We believe that, the same line of approach 
has the potential to put forward similar nontrivial results in other 
biologically relevant motifs.


\begin{acknowledgments}
We thank Alok Kumar Maity for fruitful discussion.
Financial support from Institutional Programme VI - Development of Systems 
Biology, Bose Institute, Kolkata is thankfully acknowledged.
\end{acknowledgments}


\end{document}